# Source mechanism and rupture directivity of small earthquakes in the Changning region, China, using a dense array data

Youjie Jiang[1], Jiewen Zhang[1,2], Jinping Zi[1,2], and Hongfeng Yang[1,2,3*]

1. Department of Earth and Environmental Sciences, The Chinese University of Hong Kong, Hong Kong S.A.R., China

2. Shenzhen Research Institute, The Chinese University of Hong Kong, Shenzhen, China

3. Institute of Environment, Energy and Sustainability, The Chinese University of Hong Kong, Hong Kong S.A.R., China

*Corresponding author: hyang@cuhk.edu.hk

## Declaration of competing interests

The authors acknowledge that there are no conflicts of interest recorded.

**Abstract**

Integrating focal mechanism solutions with rupture directivity analysis enables high-resolution characterization of subsurface fault geometry and earthquake rupture processes. However, resolving these parameters for small-magnitude earthquakes remains challenging due to small rupture sizes, short durations, and low signal-to-noise ratio (SNR). Here, we utilized a dense array of nodal seismometers in the Changning region, Sichuan Basin, China, to study the focal mechanism and rupture directivity of aftershocks following the 2019 Ms 6.0 induced earthquake. Using PhaseNet+ and SKHASH, we first enhance the focal mechanism catalog (1<*M*<4). Then, applying the spectral ratio method with empirical Green's functions (EGF), we observe azimuth-dependent corner frequencies of two M3 aftershocks, by spectral fitting to the Brune's model, which are consistent with unilateral rupture. Our results reveal that the two earthquakes occurred at an unidentified conjugate fault and ruptured towards N60°E unilaterally, which significantly differs from the northwestward rupture of the $M_S$ 6.0 mainshock. Furthermore, we obtain a rupture speed of approximately 0.6 times the shear wave velocity. We also apply the spectral decomposition method to compute stress drops (*M*>1), and their spatial variability reflects a long-term interplay between fluid injection and faults in the salt-mining area. These findings illuminate a complex fault system beneath the Changning anticline and highlight the importance of high-resolution seismic arrays in resolving rupture processes of small-magnitude events.

**Introduction**

Determining source parameters of earthquakes, such as focal mechanisms, rupture directivity, and stress drop, is important in understanding subsurface fault structures, rupture kinematics and dynamics, and regional seismic hazard assessment. Focal mechanisms can be applied to illuminate the geometry of active faults and regional stress field orientations (e.g., Kisslinger, 1980), rupture directivity patterns reveal the dynamic evolution of fault systems and their influences on seismic energy distribution (e.g., Somerville et al., 1997; Abercrombie et al., 2017), and stress drops provide critical information about fault mechanics and energy radiation efficiency during rupture (e.g., Boatwright, 1978; Allmann and Shearer, 2007). The detailed earthquake source characterization is particularly crucial for induced seismicity, dominated by small magnitude earthquakes while impacts profoundly in the operational safety of hydrocarbon and minerals production (Ellsworth, 2013; Yang et al., 2017). The numerous aftershocks of intermediate and large induced earthquakes provide valuable opportunities to identify and characterize the spatial and temporal seismicity patterns and the source properties (e.g. Zi et al., 2025). Such comprehensive analysis of aftershock sequences is essential for improving seismic hazard assessment and developing more effective risk mitigation strategies in industrial operations.

Intermediate and large earthquakes have the abundant low-frequency waveform information and good station coverage, benefiting for the source parameter determination. Reliable focal mechanisms for intermediate to large earthquakes can be obtained through broadband seismic waveform inversion methods such as Cut and Paste (CAP) (Zhu &

Helmberger, 1996) and ISOLated Asperities (ISOLA) (Sokos et al., 2008), while rupture directivity can be effectively characterized through azimuthal variations in source duration (Ni et al., 2005) and wave amplitude (Tan and Helmberger, 2010). However, determining source parameters for small-magnitude events is significantly challenging due to the higher frequency signals scattering by small-scale heterogeneities, limited azimuthal differences in ground motion energy, and smaller rupture areas.

Previous studies have demonstrated that rupture characterization is a critical source feature across a wide range of earthquake magnitude (e.g., Chen et al., 2021; Lin et al., 2025; Tang et al., 2024; Yang et al, 2022; Jia et al, 2025). Recent advances in seismic monitoring, particularly the deployment of dense seismic networks, have enabled more detailed studies of fault complexity (e.g., López-Comino and Cesca, 2018; Trugman et al., 2021) and rupture directivity of small earthquakes (e.g., Folesky et al., 2016; Király-Proag et al., 2019). Focal mechanisms of small events can now be determined using first-motion polarity and S/P amplitude ratio methods (Hardebeck and Shearer, 2002, 2003). In the Sichuan basin, such methods reveal that the faulting mechanisms of induced earthquakes in the Weiyuan shale gas field are primarily dominated by reverse rupturing on preexisting NNE-trending faults, with a minority of normal faulting events co-located with reverse faulting ones (Song et al., 2025). The divergence in faulting styles indicates distinct structural controls or stress conditions, highlighting the critical role of focal mechanism determination of small earthquakes in understanding local fault complexity and stress heterogeneity. On the other hand, estimating the rupture directivity of small earthquakes primarily relies on analyzing the durations and spectra of body waves (e.g., Boore and Joyner, 1978; Warren and Shearer, 2006; Cesca et al., 2011; Chen et al. 2021). Recently, using only one near-source strong motion station, the polarization of S waves also can infer the rupture directivity of small earthquakes (Yao et al., 2025). The combination of the focal mechanism inversion and directivity estimation can reduce nodal plane ambiguities for small events, especially for a complex fault network where such ambiguities cannot be resolved by earthquake locations and focal mechanism solutions.

Here we focus on the Changning region of the Sichuan Basin where massive shale gas exploration and salt mining have been conducted (Fig. 1), accompanied by gradually larger earthquakes over time in recent years (Lei et al., 2013). The June 2019 $M_S$ 6.0 Changning earthquake, probably the largest induced event by industrial exploitation in the world to date (Liu and Zahradník, 2020), was followed by four $M_L$ > 4 earthquakes and a series of aftershocks in neighboring regions, covering a span of approximately 20 km in space (Fig.1a). While the mainshock's source mechanism and directivity have been systematically investigated (e.g. Li et al., 2020; Anyiam et al., 2024), the aftershocks' rupture characteristics remain poorly constrained. If the focal mechanism and the rupture directivity were unaccounted for, it may cause biases in the measurement of earthquake source studies, such as stress drop, which are important for estimating seismic hazards and subsequent risk of industrial production near faults (Holmgren et al., 2019; Yu et al., 2022; Zhang et al., 2024).

In this study, utilizing a dense array with wide azimuthal coverage in the Changning region, we first compile an enhanced focal mechanism catalog by constraining small earthquake ($M$>1) focal mechanisms with first-motion polarity measurements, and then determine the rupture directivity for two small earthquakes (*M*~3) in detail. Applying the state-of-the-art PhaseNet+ (Zhu et al., 2025) and SKHASH (Skoumal et al., 2024), we obtain 857 high-quality solutions of focal mechanisms. Using the multiple spectral ratio method to observe the difference of azimuth-dependent corner frequencies, we investigate the rupture processes of the two small-size aftershocks ($M_L$ 3.3 and 3.2 respectively), including rupture direction, rupture speed and stress drop. We obtain stress drops for 1183 earthquakes (*M*>1) using the spectral decomposition method, and analyze the spatial distribution in the salt-mining area. These findings illuminate a complex interlocked fault system beneath the Changning anticline and highlight the importance of high-resolution seismic arrays in resolving small-magnitude earthquake rupture processes. Our results also motivate the importance of evaluating the focal mechanism and rupture directivity on the probabilistic seismic hazard and risk assessment in Changning.

## Geological Setting and Data

### Geological setting

The Sichuan Basin, situated as the most stable tectonic unit within the Yangtze craton, is bounded to the west by the Sichuan-Yunnan block. While historical seismicity activity has concentrated along its northwestern and southwestern boundaries where basement faults intersect the Longmenshan and Xianshuihe-Xiaojiang fault systems, the interior of the basin remained seismically inactive until the surging of recent industrial operations in the southern part. Since 2010, the activities of shale gas exploration and salt mining solution (dating back to the 1970s) have triggered several earthquakes with $M_L$ > 4 and shallow depths (3-5 km), such as the 2019 $M_L$ 4.9 Rongxian earthquake (Yang et al., 2020) and the 2021 $M_S$ 6.0 Luzhou earthquake (Zhao et al., 2023). In the Weiyuan shale gas field, a detailed analysis of the 2019 $M_L$ 5.6 earthquake sequence revealed 11 months of precursory fault activation directly linked to hydraulic fracturing operations (Zi et al., 2025). The Changning region, particularly the Changning salt mining area, has experienced a significant increase in seismicity rate. The $M_S$ 6.0 Changning earthquake that occurred on June 17, 2019, is the largest event in this region, leading to 13 fatalities and 299 residents injured with a total economic loss of 56.17 billion in RMB.

Our study focuses on the salt mining area in the north of the Changning region, located in the Changning-Shuanghe anticline (Fig. 1b in red part). The SE-extending asymmetric anticline features a steep northeast wing with northwest-striking faults of different dip angles, and a gentler southwest wing connecting to the wide and flat Luochang syncline (Fig. 1c). The $M_S$ 6.0 mainshock occurred in the salt mining area, and an extensive aftershock sequence depicts a complex multi-fault network of the anticlinal system, which is composed of numerous normal, reverse and strike-slip faults. Previous research (Li et al., 2020) demonstrated the mainshock is mainly a strike-slip earthquake, and both the centroid and hypocentral depths are shallow (~4 km) using the CAP method, with a unilateral rupture propagating northwestward for about 14 km. Using a local and regional waveform dataset, Anyiam et al. (2024) also determined and synthetically validated the focal mechanism solutions of the $M_S$ >5 Changning sequence and 30 $M_S$ >2 aftershocks, as well as inferred excessive pore pressures within the source zone. The aftershocks may be closely related to increased pore pressures on the seismogenic faults due to decades-long fluid injection and migration. Li et al. (2023) and Anyiam et al. (2023) suggest the seismicity is distributed close to the target shale formations in depth and along high seismic velocity boundaries in the hydraulic fracturing area. On the other hands, seismicity in the salt-mining area of the Changning region, is found mostly in areas characterized by high $V_s$ anomaly in the basement (Liu et al., 2023; Zuo et al., 2020), suggesting locally high seismic velocity and low permeability.

**Dense array**

The dense array used in this study was deployed by the Geophysical Exploration Center of the China Earthquake Administration (CEA). It consists of 298 short-period three-component portable seismometers with two sampling rates of 100 Hz and 200 Hz, starting from 6 July to 21 August 2019 in the Changning area. The seismic monitoring system consists of a square array comprising 153 stations distributed across a 60 km × 50 km region, and a linear array of 145 stations in the ranges of 27.8°-28.8°N and 104.3°-105.2°E. We only choose the stations with a sampling rate of 200 Hz (Fig. 1a) for analysis. The wide azimuthal coverage is expected to provide better resolution of small-magnitude events and their source characteristics.

**Methods**

**Focal mechanism inversion**

The robust foundation of our analysis is based on the relocated earthquake catalog (Zhao et al., 2023), which identified events ranging from $M_L$ -0.5 to 3.9 using the same dense array dataset, with 14 larger events reaching up to $M_L$ 3.0. Due to the signal-to-noise ratio (SNR) of data, we here focus on events with $M_L > 1$. Based on the previous phase arrival time of the earthquake catalog, we use PhaseNet+, a multi-task deep learning model built upon PhaseNet (Zhu & Beroza, 2019), to simultaneously predict phase arrivals and onset polarities. The predicted polarity ranges between -1 and 1 representing the probability of downward (negative) or upward (positive) onset. For example, the azimuthal P-wave polarity pattern of Event 1 (2019071410171814) reveals a left-lateral strike-slip focal mechanism (Fig. 2). Stations in the NE and SW quadrants, such as 3302 (azimuth 345°) and 3808 (azimuth 162°), exhibit dominant negative polarities, consistent with compressional quadrant radiation predicted by the strike-slip focal mechanism. Conversely, NW and SE quadrant stations, for instance, 1059 (azimuth 69°) and 3204 (azimuth 294°) display positive polarities, aligning with dilatational quadrant expectations.

To resolve the fault geometry and kinematics of the analyzed events ($M_L > 1$), we determined earthquake source mechanisms using SKHASH, largely based on the HASH algorithm (Hardebeck & Shearer, 2003). The wide coverage of our seismic network enabled us directly to utilize the predicted P-wave polarities as input data from multiple stations across diverse azimuths. The minimum number of required polarities is 8, within a maximum azimuthal gap of 90°. SKHASH implements a rigorous quality assessment framework for focal mechanism solutions through systematic misfit quantification. Solutions are assigned quality grades from 'A' to 'D', based on nodal plane constraint criteria, where Grade A requires the probability that focal mechanism solution should be larger than 0.8 (Fig. S1).

**Rupture direction**

To determine their fault planes, we estimate the rupture directivities of two relatively large events (*M*~3), 2019071410171814 (Event 1) and 2019072208190033 (Event 2), which occurred in the central section of the Changning anticline (Fig. 3 and 4). Rupture directivity can clarify faulting systems and the regional stress environment, especially for complex fault networks with multiple active fault strands.

Assuming an omega-square source model (Aki, 1967), the displacement spectrum exhibits an azimuth-dependent corner frequency that is inversely proportional to the source duration. To identify the corner frequency of target events, we apply the empirical Green's function (EGF) technique that uses the waveforms of closely located smaller events as a reference to remove common path and site effects from a target event. To mitigate the influence of biased frequency contents from S waves or coda by using a time window with a fixed length, we exclusively analyze spectrum between P- and S-wave arrivals (e.g., Fig. 5c). Furthermore, we set a cross-correlation coefficient threshold of 0.7 for the band-pass filtered waveforms inside 1-10 Hz (e.g., Fig. 5d). The selected EGFs have similar focal mechanisms to the main event and are at least one magnitude unit smaller (Fig. 5a), satisfying the EGF criteria (Abercrombie et al., 2017). Due to the limitation of instrument response, we exclude frequency contents outside 1–80 Hz range. To mitigate the influence of noise, we apply strict criteria to screen out noisy waveforms: for each spectrum, 90 percent of the frequency sampling points between 1 and 80 Hz shall have an SNR larger than 3, with the noise spectrum calculated in a 1-s window before the P-wave arrival, following the criteria as Zhang et al., (2024). The P-wave window starts 0.1 s before the direct P wave arrival and lasts to the S wave arrival. Before calculating the spectrum, we use a multi-taper method (Python package: spectrum, pmtm function) to get more stable spectral results.

The target events are Event 1 and Event 2, which occurred in the center of the Changning anticline and showed different nodal plane patterns compared with those on the two sides of the anticline (Fig. 3a). The two events are found to have 45 and 69 EGFs, respectively. Then, the corner frequency of the mainshock at different stations can be obtained by fitting the spectral ratio from the Brune spectral model (Brune, 1970) to the spectral ratio of the target and EGF events (e.g., Abercrombie, 2014; Huang et al., 2016):

$$\frac{u_1(f)}{u_2(f)} = \frac{M_1}{M_2} \frac{1+(\frac{f}{f_{c2}})^2}{1+(\frac{f}{f_{c1}})^2} \qquad (1)$$

where $M_1$ is the seismic moment of the target event, $f_{c1}$ is the corner frequency of the target event, $M_2$ is the seismic moment of the EGF event, $f_{c2}$ is the corner frequency of the EGF event.

The shape of the spectral ratios computed from different EGF events reflects the characteristics of the source spectrum for the target event. To improve the accuracy of the source spectrum, we perform multiple spectral ratio analysis, stacking spectral ratios from at least 3 high-quality EGF pairs per station. The analysis has been designed to prioritize the appropriateness of the EGF event selection, and its stability by stacking the best EGF events.

Many previous studies have characterized rupture directivity patterns as unilateral,

bilateral, or complex (e.g., Abercrombie et al., 2017; Chu et al., 2025). To quantify the directivities of two target events, we analyze the azimuthal pattern of corner frequency, which is inversely proportional to the apparent pulse duration. A unilateral rupture earthquake produces higher corner frequency and shorter pulse durations in the direction of rupture propagation. For a unilateral rupture, the apparent pulse duration $\delta t$ can be written as:

$$\delta t = t_0 - \frac{\Delta \cos(\theta)}{c} = t_0 + \Delta\Gamma \tag{2}$$

$$\Gamma = \frac{-\cos(\theta)}{c} \tag{3}$$

in which $t_0$ and $\Delta$ are the true time difference and rupture distance, respectively. $\theta$ is the azimuth of the station relative to the rupture direction, $c$ is the considered phase velocity (P wave in our study) near the source, and $\Gamma$ is the directivity parameter.

**Rupture speed**

The good azimuthal coverage of the dense array also enables simultaneous resolution of fault geometry, rupture length and rupture speed. Rupture velocity is a key source parameter to describe the dynamic property of seismic source. Assuming unilateral directivity in a horizontal plane (e.g., Boatwright, 2007), the directivity coefficient through the corner frequency ratio is:

$$Cd(\theta) = \frac{f_c}{f_a} \tag{4}$$

$$Cd(\theta) = \frac{1}{1-\left(\frac{v_r}{v_s}\right)\cos(\theta-\theta_0)} \tag{5}$$

in which $f_a$ is the apparent corner frequency, $f_c$ is the true corner frequency as a function of azimuth ($\theta$), $v_r$ and $v_s$ are the rupture velocity and shear-wave speed, respectively, $\theta$ is the station azimuth and $\theta_0$ is the rupture direction.

Based on the rupture direction method above, we can get the rupture direction $\theta_0$ and the apparent corner frequency. Combining with Equation (4) and (5), it is able for us to obtain the best fitting of the ratio of the rupture velocity and shear-wave speed.

**Stress drop estimation**

Using the spectral ratio method, we can only estimate stress drops for relatively large events (*M*~3). Because the EGFs are at least one magnitude unit smaller than the main event (Abercrombie et al., 2017). To estimate stress drops for more small earthquakes, we apply an improved spectral decomposition method (Zhang et al., 2024; Zhang et al., 2025). Using a global-optimization algorithm, the Differential Evolution (DE) (Storn and Price, 1997), this new approach reduces the uncertainty of the corner frequency estimation and finds the optimized combination of stress drops in all the magnitude bins. The observed spectra can be analyzed as the convolution of an event term, a path term, and a site term. More specifically, for earthquakes recorded by at least 3 stations, we use the spectral decomposition method to solve for their event spectra (Shearer et al., 2006), and obtain a correction term, the Empirical Correction Spectrum (ECS), using the DE method to isolate the source spectra from the event spectra. Then, the earthquake corner frequencies are estimated by fitting the source spectra to the Brune's source spectrum model:

$$u(f) = \frac{\Omega_0}{1+(\frac{f}{f_c})^2} \qquad (6)$$

in which u(f) is the source spectrum, $\Omega_0$ is the zero-frequency spectral level (proportional to log $M_0$), and $f_c$ is the corner frequency.

We obtain 1475 reliable P-wave event spectra from the 1989 recorded events (1< $M_L$ <4). To obtain the moment magnitudes of these earthquakes, we follow the magnitude calibration workflow as Shearer et al. (2006), Zhang et al. (2022) and Vandevert et al. (2025), to establish a conversion relationship from local magnitude to moment magnitude, utilizing the average amplitude in the low-frequency plateau of the event spectra. We average the event spectra within bins of 0.2 units in local magnitude. There are only 7, 3, 3 spectra contributing to the bin 11, 12 and 13, separately, compared to 155 for $M_L$ = 1.6-1.8 and 222 for $M_L$ = 1.4-1.6 (Fig. S2b). The bins with few data (<10 events) have irregular shaped spectra compared with other bins stacking a larger number of spectra. To improve the estimation of ECS (Shearer et al., 2006; Shearer et al., 2019), we use the $M_L$ = 1.0 to 3.0 results to calculate the final ECS (Fig. S10), and then correct their event spectra ($M_L$>1).

We compute the median estimated log $M_0$ for $M_L$ bins of width 0.2 magnitude to better fit the line (Fig. 6). We convert the relative log moments to absolute moment estimates ($M_0$) and their corresponding $M_W$ estimates (Hanks and Kanamori, 1979). Based on the scaling of the log $M_0$ and the local magnitude for P waves, the linear relationship of $M_L$ and $M_W$ in the study region is

$$\begin{cases} M_W = 0.68 \times M_L + 0.96, 1 < M_L < 3 \\ M_W = 0.84 \times M_L + 0.48, M_L > 3 \end{cases} \quad (7)$$

From the estimated $f_c$ and $M_0$ values, we estimate Brune stress drops (Brune, 1970) as,

$$\Delta\sigma = \frac{7}{16}\left(\frac{M_0}{r^3}\right) = \frac{7}{16} M_0 \left(\frac{f_c}{kV_s}\right)^3 \quad (8)$$

$$f_c = k\frac{V_s}{r} \qquad (9)$$

where $V_s$ represents shear velocity, $M_0$ is the seismic moment of the earthquake, $k$ is a constant depending on model assumptions, and $r$ stands for source radius.

For the shear velocity, we use the tomography results of the Changning region from Li et al. (2023) (Fig. S3). The average 1D velocity model correlates with low Vs in shallow crust. Beneath the Changning anticlinal system, the Cambrian to Silurian caprocks overlying the Sinian Dengying Formation are target layers for salt extraction (Jia et al., 2006; Anyiam et al., 2024). Specially, the Dengying Formation, an overlying quartz- and silica-rich slip-prone caprock, is between 2.7 and 3 km depth (Sun et al., 2017; Lei et al., 2019).

## Results

Because of the dense array's wide azimuth coverage, we achieve robust solutions only using P-wave polarities, resulting in a lower uncertainty (Fig. S1). The resulting focal mechanism catalog has 1792 solutions of varying qualities, accounting for ~87% of earthquakes (*M* > 1) from the hypocenter catalog. We focus on 857 focal mechanisms associated with the highest quality, 'A' (Fig. 3a). In particular, the 'A' quality focal mechanisms have a larger number of polarity picks ($\geq 24$), lower fault plane uncertainties (<30°), higher probabilities (>80%), and lower polarity misfits (<15%) (Fig. S4).

Based on the selected high-quality focal mechanisms, we further identify the principal axes (P, B, T) to characterize the potential faulting types according to the Kaverina projection (Álvarez-Gómez, 2019). Our results reveal the compressional types of faulting, including Strike-slip (39%), Strike-slip-Reverse (33%) and Reverse-Strike-slip (19%) (Fig. 3b). There patterns are consistent with both aftershock focal mechanisms in the salt-mining region (Yi et al., 2019) and the regional tectonic background. The prevailing orientation of maximum horizontal principal stress (SH$_{max}$) falls within the range between N30°E and N70°E on the Changning anticline (Liu et al., 2021), which aligns with the P-axis orientations from the focal mechanism solutions. The predominant faulting type for the central section is pure Strike-slip, compared with Reverse and Reverse-strike-slip types dominant in the NW and SE sections (Fig. S5).

The fitting results of the two target events, located in the center section, demonstrates a cosine relationship of azimuth-dependent corner frequencies varying across different azimuths, showing clear unilateral rupture directivity. Consequently, the minimum $1/f_c$ is the rupture direction: 65° for Event 1 (Fig. 7), 59° for Event 2 (Fig. 8). For Event 1, the corner frequency is ~16.67 Hz in the forward direction and ~4.55 Hz in the backward direction (Fig. 7a). The apparent corner frequency is 7.14 Hz, smaller than the geometric mean (9.52 Hz) and the median (8.91 Hz). For Event 2, the apparent corner frequency is 5.88 Hz (Fig. 8a). The northeast rupture directions (65° for Event 1, 59° for Event 2) align with one of the focal-mechanism nodal-plane, but notably differ from the mainshock's northwestward rupture direction. Based on the detailed analysis of the rupture directions and the apparent corner frequencies, we obtain the best fitting of the ratio of the rupture velocity and shear velocity ($v_r/v_s = 0.64$) for Event 1, when the shear velocity is 3.2 km/s (Fig. 7e). Meanwhile, for Event 2, $v_r/v_s = 0.59$ (Fig. 8e).

For equation (8), assuming asymmetrical circular source, $k = 0.21$ for P-wave when $v_r/v_s = 0.6$ (Kaneko and Shearer, 2014). Here, we use the apparent corner frequency obtained from the fitting result and the shear velocity from Li et al. (2023). The stress drops are 29 MPa for Event 1, and 12 MPa for Event 2, respectively. In addition, in spectral decomposition method, we use $k = 0.38$ for the P-wave results (Kaneko and Shearer, 2014) and the same shear velocity model (Li et al., 2023) to estimate stress drop. We obtain stress drops for a total of 631 events ($M_L$>1) within the salt-mining area (Fig. 9). The median and geometric mean of stress drops are 7.91 and 8.83 MPa, respectively. We identify that the center section displays elevated median stress drops (Δσ = 9.36 MPa) compared to the northeastern and southwestern sections (Δσ = 6.13 and 8.90 MPa, respectably) (Fig. 9 and S6b). The stress drops are observed significantly increasing in the depth range of 2–4 km, but their depth dependence disappears below 6 km (Fig. 9c). Meanwhile, the stress drops are found not visibly scaled with moment magnitudes (Fig. 9d), supporting the self-similarity assumption (Aki, 1967). A few of the smallest events deviate from this trend due to the estimated corner frequencies outside the frequency band.

**Discussion**

**Uncertainty of small earthquake focal mechanisms and rupture directivities**

The SKHASH algorithm uses a grid-search approach that identifies a set of acceptable focal mechanism solutions (Skoumal et al., 2024). In our study, we set the fraction of assumed bad polarities at 0.1. The acceptable solutions are defined as those with a P-wave polarity misfit below this threshold. SKHASH then determines the best-fit focal mechanism by averaging these acceptable solutions, given the possible source depth and velocity model. The quality rankings of focal mechanism are based on the statistical tightness of the acceptable solution set and the number of polarity misfits for the preferred solution. Meanwhile, the uncertainty is determined by the observed polarities and computed takeoff angles, as well as the RMS difference from the preferred solution and acceptable solutions.

The uncertainty in focal mechanisms solutions for small earthquakes is mainly influenced by the quantity of P-wave polarities and the station distribution (Yang et al., 2012). The reliability of PhaseNet+ polarity identification has been examined by Song et al. (2025). Due to the wide station coverage, we achieve reliable solutions only using P-wave polarity, including a lower uncertainty (Fig. S1). Notably, the nodal plane uncertainty of all qualities focal mechanisms correlates with the focal mechanism probability (Fig. S4c) and the polarity misfit (Fig. S4d). Such trends with the magnitude (Fig. S4a) and the number of polarities ((Fig. S4b) are less obvious. It remains challenging to derive reliable focal mechanism solutions for smaller earthquakes ($M$<1.5) given the station coverage. Therefore, in this study we only account for the good quality 'A' as our results to mitigate the impact of uncertainty in focal mechanism inversion. The 'A' quality focal mechanism solutions show consistency with the regional stress field observations (Zhang et al., 2022) in the three distinct seismotectonic subregions.

Station coverage is the primary reason that affects our analysis of rupture directivity. We apply a bootstrap analysis, performing 1,000 bootstrap iterations to evaluate the rupture azimuth uncertainty. It reveals that rupture azimuth uncertainties exceed 10° with fewer than 12 recording stations for Event 1 (Fig. S7), and 18 stations for Event 2 (Fig. S8).

The rupture directivity estimates align with one of the nodal planes (e.g., Fig. S9), confirming the robustness of the resolved focal mechanism solution and rupture direction.

**Analysis of stress drop estimation**

The stress drops for Events 1 and 2, determined as 29 MPa and 12 MPa respectively using the spectral ratio method, are significantly higher than the median stress drop of 9.36 MPa estimated for the central Changning anticline segment using the spectral decomposition method. However, due to the lack of events for $M$>3 which could result in unreliable stress drop estimates in this magnitude range (Fig. S2), we prefer to compare the corner frequency results rather than the absolute stress drop values. The corner frequencies obtained from the spectral decomposition method are 9.31 and 8.94 Hz for Events 1 and 2, higher than the apparent corner frequencies considering rupture directivity using the spectral ratio method (Tab. S1). Specifically, for Event 1 which shows a better fit in the rupture direction analysis, the corner frequency from the spectral decomposition method aligns more closely with the median corner frequency determined by the spectral ratio method. This could be understood as that the spectral-decomposition-based method estimates a corner frequency for each event by averaging the spectra across all azimuths, which should approximate the median corner frequency of the same event from the spectral ratio method. However, if the directivity effect is observed for an event, then the corner frequency estimates accounting for the directivity effect should provide a more accurate observation for earthquake rupture processes, than simply calculating a median corner frequency across different stations.

While the stress drop estimation of absolute values can be uncertain, the relative stress drop measurements are much more reliable (Shearer et al., 2019; Shearer et al., 2024). Our analysis focuses on the results obtained from the spectral decomposition method within events where $M_L$ > 1 (Fig. S10). The analysis of depth and magnitude distributions (Fig. S6 and S11) shows a median aftershock depth of 4 km. A higher frequency of earthquakes occurs at depths of approximately 3-4 km and moment magnitudes of 1.8-2.0, particularly near the high pore pressure saltwater injection layer (~3 km) (Yang et al., 2024). The central section shows deeper seismicity (~5 km), potentially associated with a conjugate fault (Fig. 9b). The NW section exhibits the deepest seismicity (5–8 km) with larger magnitudes along the east-west-trending anticlinal axis (Fig. S6d). The SE section, situated near the mainshock and salt-mining injection wells, features lowest stress drop compared with the other two sections. The lower stress drops in the SE section likely reflect regional stress relaxation and reduced stress accumulation resulting from long-term saltwater injection operations at relatively shallower depths over three decades, while the relatively higher stress drops in the NW and central sections are more associated with generally deeper earthquakes distant from injection spots.

The minimized stress drop scaling with moment magnitude (Fig. 9d) may indicate earthquake self-similarity within the three sections of the salt-mining area. The stress drop scaling with depth (Fig. 9c) shows a significant increase in stress drop within the depth range of 2–4 km. The Cambrian to Silurian caprocks within this depth range (Fig. S3) are extensively fractured and faulted (Anyiam et al., 2024). Influenced by salt-water injection, the high pore pressure in the target salt extraction formation reduces the effective normal stress, thereby facilitating slip on fractures and faults that might otherwise have remained stable. The prolonged injection unloads shear stress at the depth of ~3 km, resulting in lower stress drops and an increased frequency of earthquakes. It is challenging to apply reasonable corrections for depth-dependent attenuation and rupture velocity to eliminate the potential systematic dependence of stress drop on source depth within the upper 6 km of the salt-mining area. The wider depth range observed in the northwest section (Fig. S6a) primarily influences the stress drop scaling with depth.

**Heterogeneity of the Changning anticline**

The primary rupture plane of the mainshock is a NW-SE striking fault (strike 125°; Li et al., 2020). However, the diversity of nodal plane orientations for the aftershock sequence, evidenced by the wide distribution of P-, T- and B-axis orientations reveals significant complexity in fault geometry and stress patterns. Particularly, T-axis plunges are predominantly between 0° and 70° (Fig. 3b). Meanwhile, the resolved SE rupture directivity (60°±10°) for several aftershocks is highly oblique to the mainshock's fault plane, indicating the activation of conjugate fault structures in the central anticline axis (Fig. 4). Dai et al. (2023) identified that on 23 Jun 2019, a $M_W$ 4.3 aftershock occurred near the identified conjugate fault, was different from other $M_W$ > 4 aftershocks. The $M_W$ 4.3 event shows shear stress and high fluid overpressure, likely caused by the relatively large difference between the fault strike and the direction of the tectonic stress field for the mainshock.

The two $M$~3 events exhibit notably slow rupture speeds (Vr/Vs ~ 0.6), approaching the lower bound of the typical patterns where Vr/Vs ranges from 0.6 to 0.8 (Heaton, 1990) and notably slower than the average of 0.72 for moderate to large earthquakes (Geller, 1976). While small earthquakes (3.5 ≤ $M$ ≤ 4.1) with strong directivity typically show faster rupture velocities (0.63Vs ≤ Vr ≤ 0.87Vs), events with weak directivity propagate more slowly or bilaterally (Boatwright, 2007). The cosine relationship of azimuth-dependent corner frequencies demonstrates unilateral rupture directivity for the two target events. Smaller earthquakes are known to exhibit higher sensitivity to local geological conditions (Abercrombie et al., 2017; Huang et al., 2020). Such slow rupture velocities likely result from a combination of fault geometry complexity and high pore pressure from prolonged salt-mining operations.

The 17 June 2019, $M_S$ 6.0 Changning earthquake aftershocks provides a significant opportunity to investigate how a large mainshock interacts with a complex, pre-existing fault system under the influence of industrial activities. Our analysis of the aftershocks reveals that their rupture processes are not simple repetitions of the mainshock but instead map out a conjugate fault architecture and a locally perturbed stress field. The directivity observations for the central Changning fault section qualitatively agree with the focal mechanism patterns. Combined with the stress-drop variations, it can provide robust insights into fault interaction mechanisms. Despite over 30 years of saltwater injection, most aftershocks show strike-slip faulting type, consistent with the regional stress patterns. It also suggests that induced seismicity following large earthquakes has lower seismic risks in our study area, unlike the more abrupt stress perturbations observed in hydraulic fracturing zones of the Changning region (Dai et al., 2025). Therefore, analyzing source mechanisms and rupture directivities in these smaller events can help forecast regional fault systems, stress distribution, and the potential locations and characteristics of future induced earthquakes. These findings enhance our understanding of smaller induced seismicity processes and contribute to improved seismic hazard assessment in regions affected by energy production operations.

**Conclusion**

In this study, we first enhance the focal mechanism catalog of 857 earthquakes with $M_L$>1 occurring in the salt-mining area of the Changning region from 6 July to 21 August 2019. Our high-resolution results show the strike-slip rupturing of aftershocks in the Changning anticline is the primary faulting pattern, and the spatial diversity of nodal planes may result from fault geometry complexity or stress field heterogeneity at short-length scales. We focus on the middle segment of the Changning anticline and comprehensively estimate the rupture process of two induced earthquakes ($M \sim 3.0$) by applying the multiple EGF approach. It indicates northeastward rupture propagation (~60°) along a previously unidentified strike-slip conjugate fault at a relatively low rupture velocity (~0.6Vs), contrasting with the northwestward rupture of the $M_S$ 6.0 mainshock. Using the spectral decomposition method, we observe a spatial distribution of stress drop for three subsections in the Changning anticline. The stress drops are significantly increasing in the depth range of 2–4 km, which corresponds to the salt extraction layers. These findings contribute to our understanding of rupture processes across different earthquake magnitudes and highlight the importance of high-resolution seismic arrays in identifying source parameters of small-magnitude seismic events.

## Data and Resources

Determination of pick polarities is conducted using the deep learning model PhaseNet+ (Zhu et al., 2025; https://github.com/AI4EPS/EQNet), and the earthquake catalogue is produced by Zhao et al (2023b) (DOI: 10.1016/j.tecto.2023.230086). The focal mechanism catalog is available from https://zenodo.org/records/15450041. Data processing largely depends on ObsPy (Beyreuther et al., 2010; https://docs.obspy.org). Figures are made using PyGMT (https://www.pygmt.org) and Matplotlib (https://matplotlib.org). The supplemental material includes additional figures that provide more detailed insights into the results and discussion.

## Declaration of competing interests

The author acknowledges that there are no conflicts of interest recorded.

## Acknowledgments

We thank the Geophysical Exploration Center of China Earthquake Administration for providing the seismic waveform data. This study was supported by the National Natural Science Foundation of China (No. U2139203), Hong Kong Research Grant Council Grants (No. 14303721, 14306122), Shenzhen Fundamental Research Program (Shenzhen Natural Science Foundation, No. JCYJ20250604141405007) and the CUHK Vice Chancellor's PhD Scholarship Scheme.

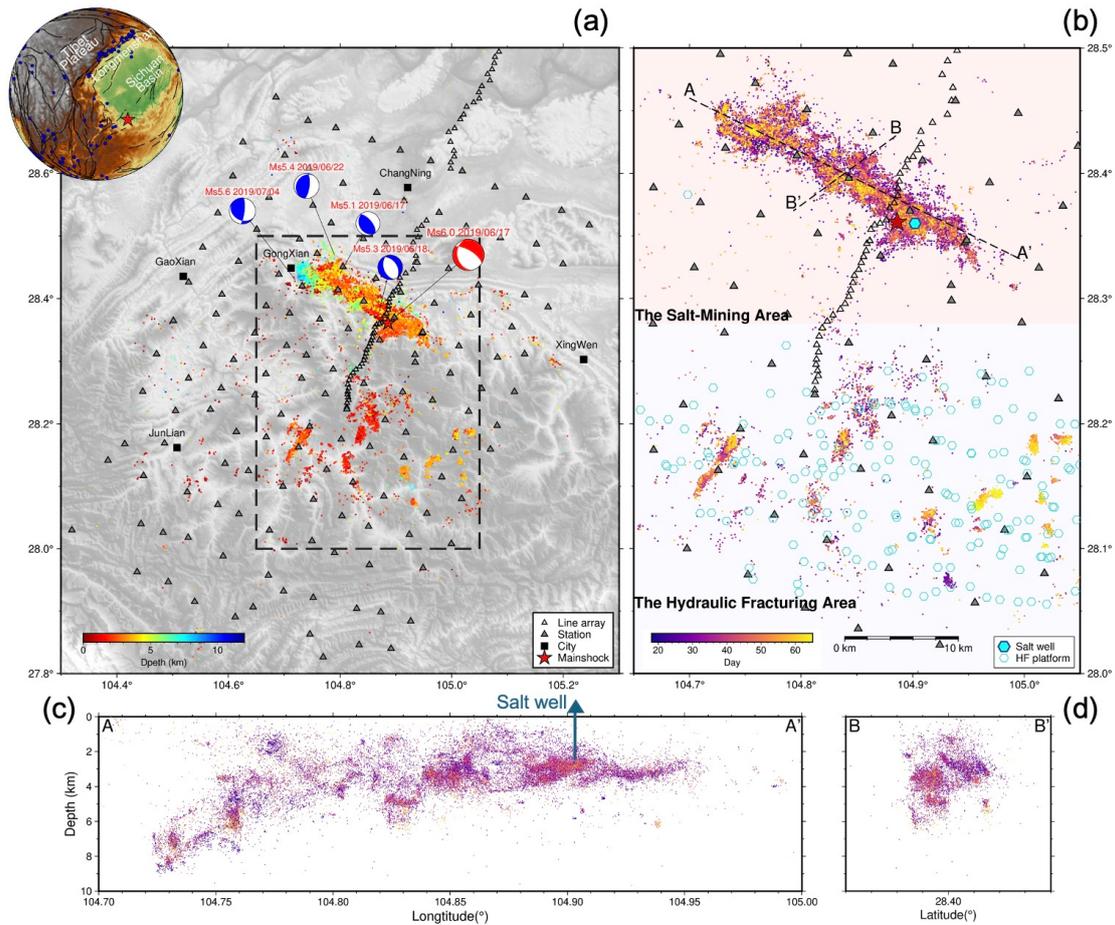

Figure 1. (a) Relocated seismicity using utilized dense array in Changning region. Beachballs (Anyiam et al., 2023) represent focal mechanisms of the $M_L$ >5 Changning earthquake sequence. The inset shows locations of historical (1972-2015) $M_L$ > 5 earthquakes in blue dots (https://data.earthquake.cn) in and around the Sichuan Basin. (b) Zoomed in the view of the dashed box in (a). (c) and (d) Cross-section views of aftershocks for A - A' and B - B' in (b).

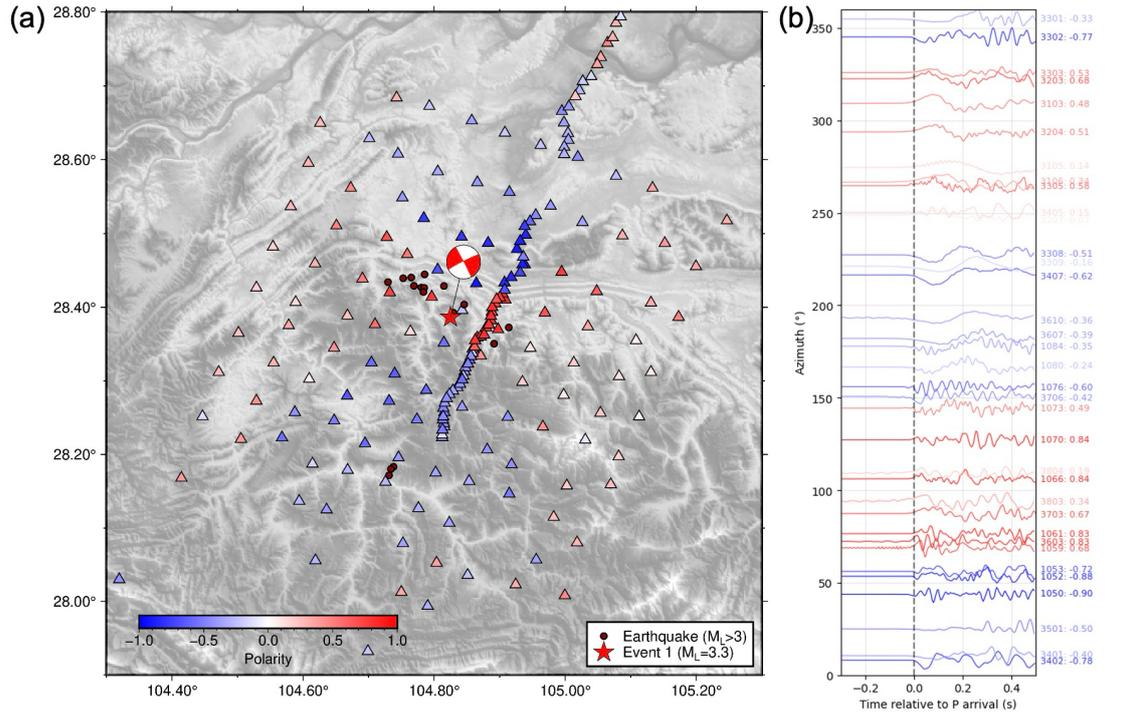

Figure 2. (a) Focal mechanism inversion of Event 1 using P wave polarities. (b) Waveforms within the 0.8-s window (0.3 s before the P arrival and 0.5 s after). The dashed line represents the predicted P arrivals for Event 1. Red and blue indicate positive and negative first motion polarity prediction as (a), followed by the station name and the probability or the confidence.

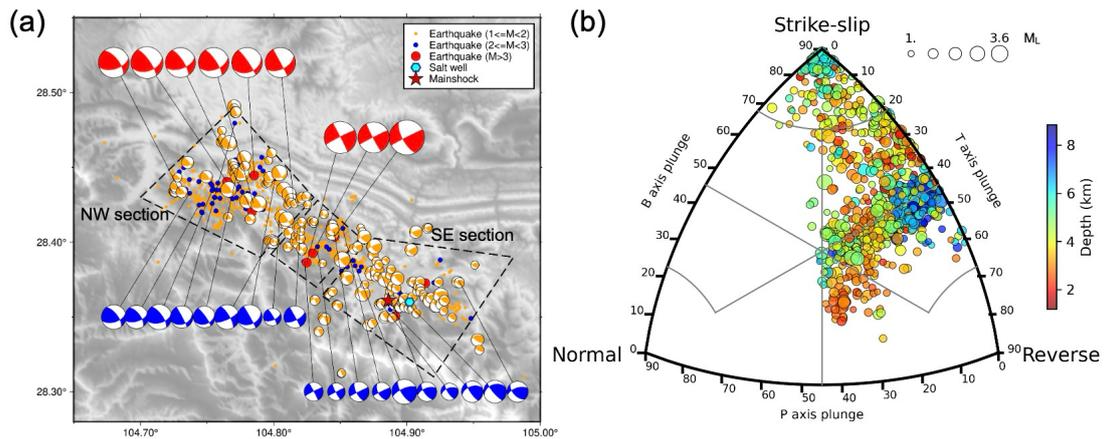

Figure 3. (a) Focal mechanisms of earthquakes ($M_L$ >1) in the salt mining area using SKHASH. (b) Faulting types of focal mechanisms determined from (a).

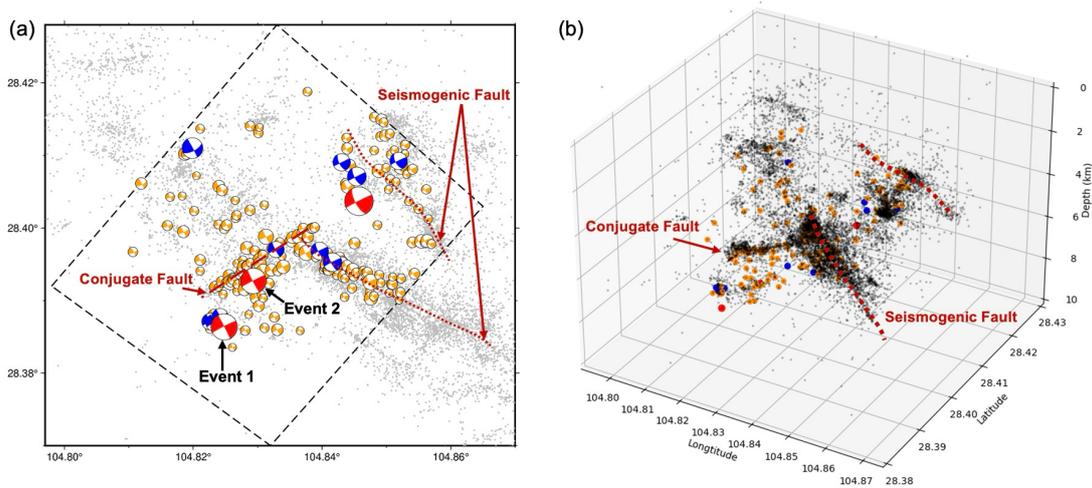

Figure 4. Focal mechanisms of earthquakes in the center section of the Changning-anticline fault system, and inferred fault architecture from seismicity, in horizontal map (a) and 3D (b). The inferred faults as shown by dashed red lines align well with the aftershock locations (gray dots). The colors of beach ball are same as the Fig. 3.

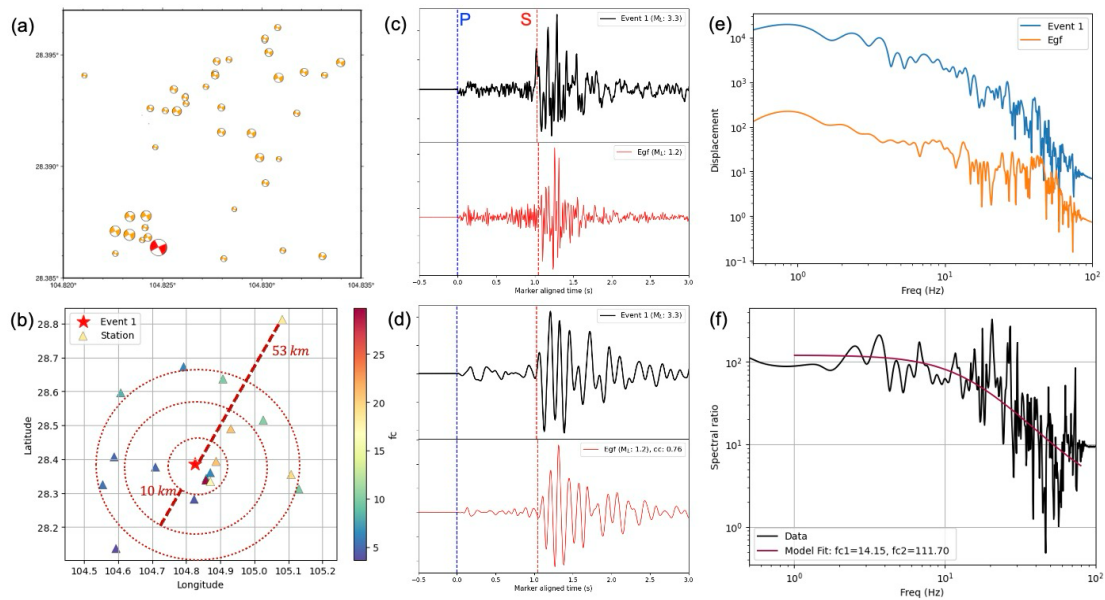

Figure 5. (a) A map shows the locations and focal mechanisms of Event 1 (red) and its EGFs (yellow). (b) Spatial distribution of fc at different stations. (c) – (f) One pair with Event 1 and its EGF. (c) The raw vertical-component waveform. (d) Normalized filtered waveform (1 - 10 hz). (e) and (f) The P wave displacement spectra and the spectral ratio of this pair.

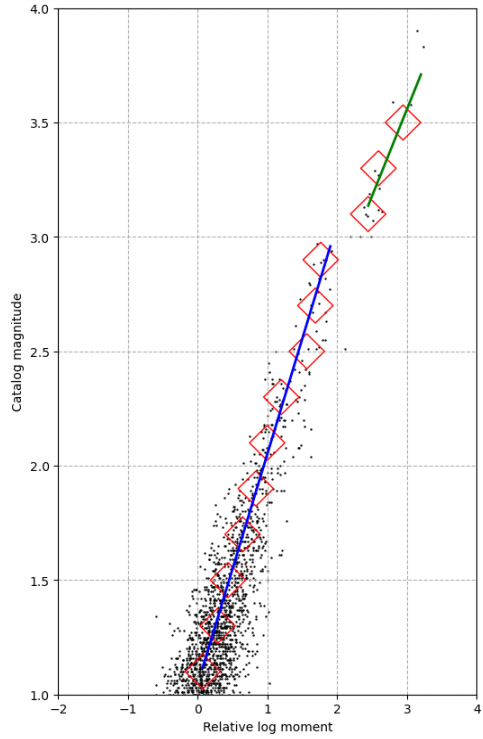

Figure 6. Catalog magnitude ($M_L$) versus relative log moment, binned in increments of 0.2 in $M_L$ (red square), and the best-fitting lines to the resulting bins are shown for 1< $M_L$ <3 (in blue) and $M_L$ >3 (in green).

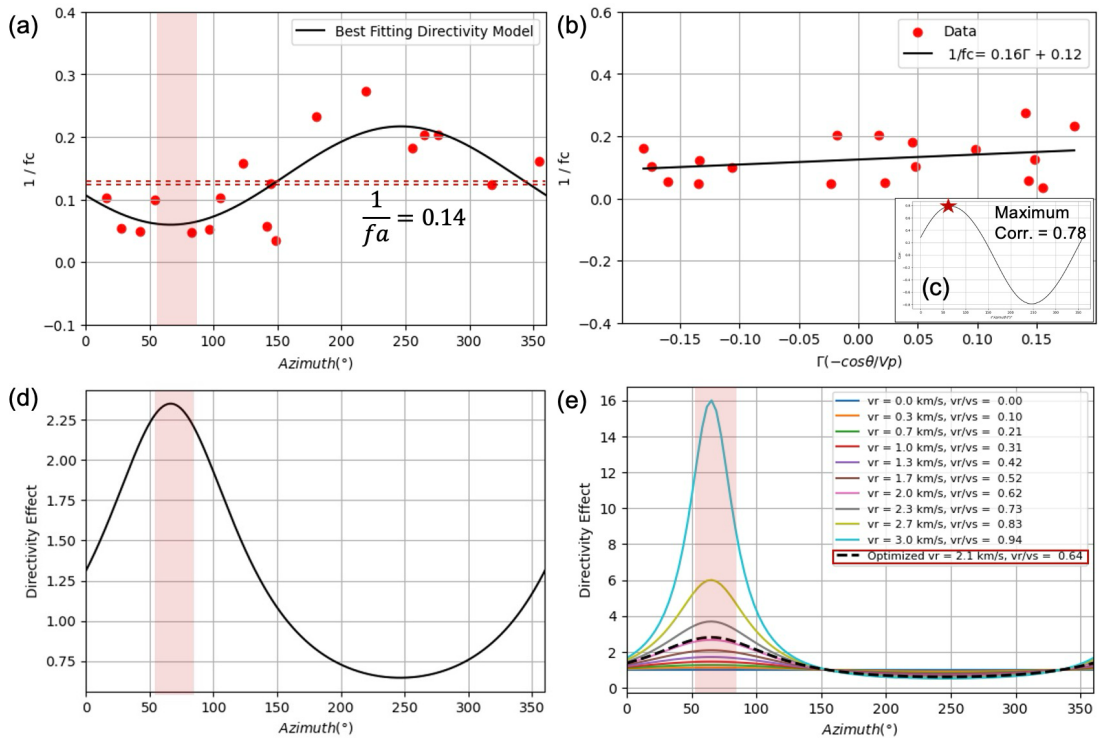

Figure 7. Source parameters analysis of the Event 1. The black solid lines in (a) and (b) show the best fitting directivity model. (c) The correlation coefficient (Corr.) as a function of possible azimuths, with the optimal rupture direction marked by a star. (d) The directivity effect of the Event 1. The dash lines in (e) shows the best fitting rupture speed verse the shear velocity.

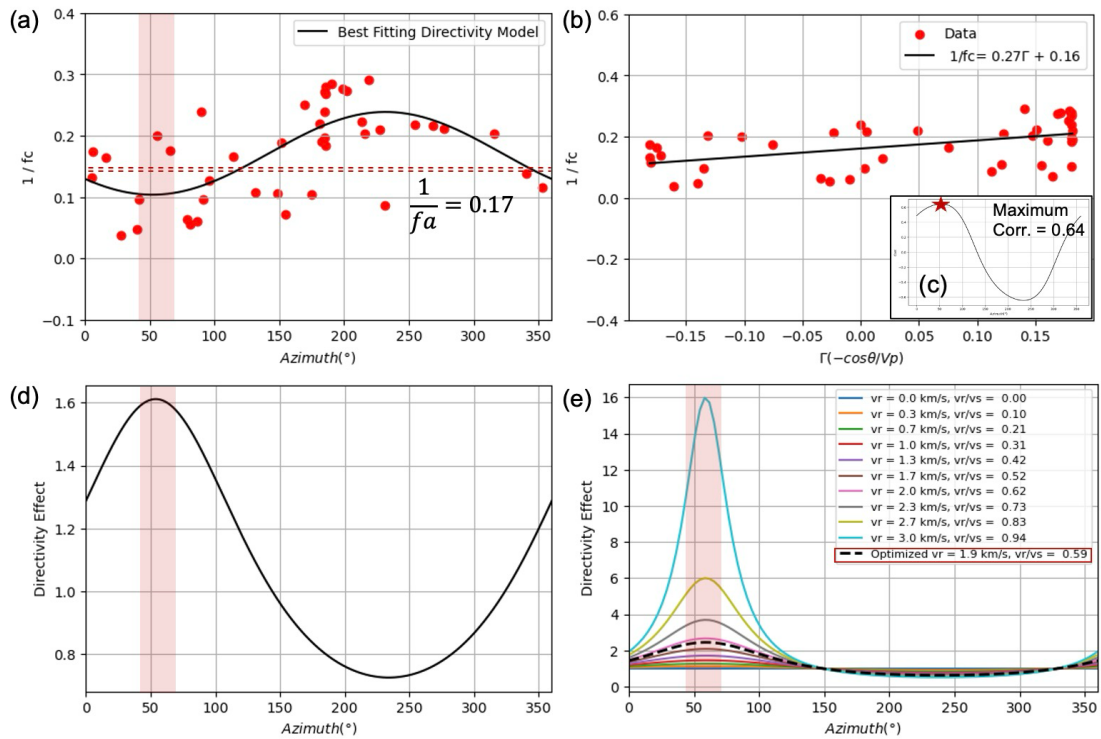

Figure 8. Source parameters analysis of the Event 2. The black solid lines in (a) and (b) show the best fitting directivity model. (c) The correlation coefficient (Corr.) as a function of possible azimuths, with the optimal rupture direction marked by a star. (d) The directivity effect of the Event 2. The dash lines in (e) shows the best fitting rupture speed verse the shear velocity.

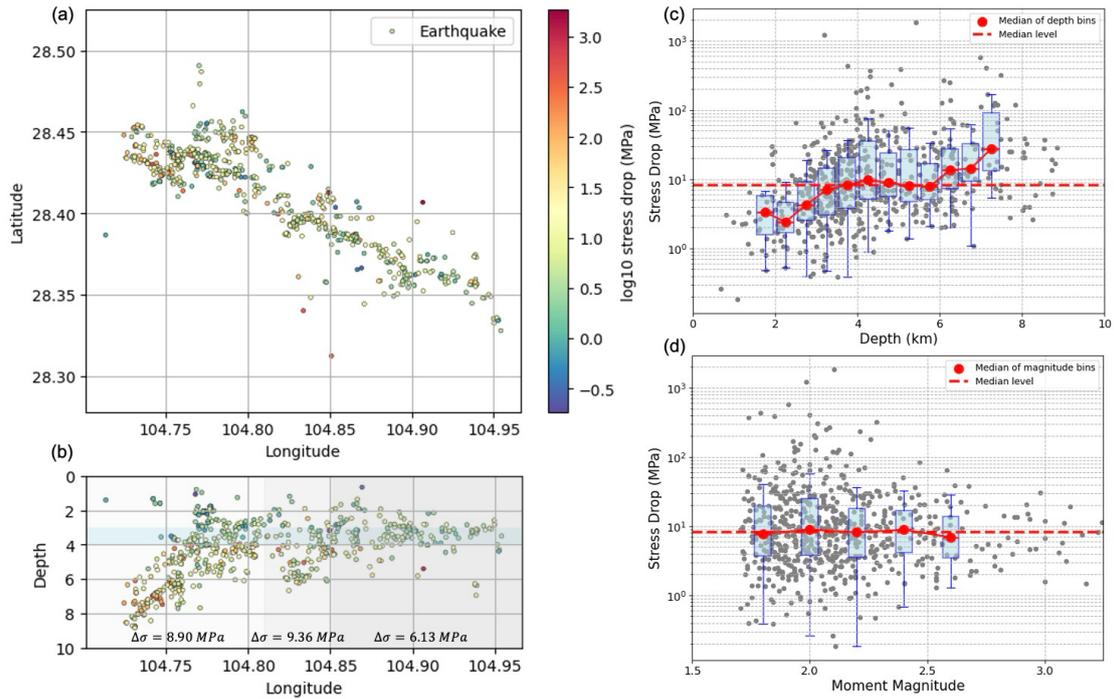

Figure 9. (a) and (b) Stress drop estimates for *M* 1.0 to 4.0 earthquakes in the salt-mining region, Changning. Three sections of the Changning anticline fault system are showed by same color as Fig. 3c. The stress drops are the median value of each section, and the light blue area shows the salt water injection layer. (c) and (d) stress drop dependence on depth and magnitude, respectively. The blue dots represent the individual stress drop values, and the red dots represent the median values in moment magnitude bins with an interval of 0.2, and depth bins with an interval of 0.5 km. It is required that in each bin there are at least 20 data points. The rectangle boxs represent the 25th (lower bar) and 75th (higher bar) percentiles of the values.The error bars extend from the box to the farthest data point lying within 1.5x the inter-quartile range from the box. The horizontal red dashed lines indicate the median stress drop levels in these cases.